# Hollow-core fiber-based speckle displacement sensor

Jonas H. Osório,[1,*] Thiago D. Cabral,[1, 2] Eric Fujiwara,[2] Marcos A. R. Franco,[3] Foued Amrani,[4] Frédéric Delahaye,[4] Frédéric Gérôme,[4] Fetah Benabid,[4] Cristiano M. B. Cordeiro[1]

[1]*Institute of Physics Gleb Wataghin, University of Campinas, Campinas, Brazil*
[2]*School of Mechanical Engineering, University of Campinas, Campinas, Brazil*
[3]*Institute for Advanced Studies, IEAv, São José dos Campos, Brazil*
[4]*GPPMM group, XLIM Institute, CNRS UMR 7252, University of Limoges, France*
*\*Corresponding author: jhosorio@ifi.unicamp.br*

**The research enterprise towards achieving high-performance hollow-core photonic crystal fibers has led to impressive advancements in the latest years. Indeed, using this family of fibers becomes nowadays an overarching strategy for building a multitude of optical systems ranging from beam delivery devices to optical sources and sensors. In most applications, an effective single-mode operation is desired and, as such, the fiber microstructure or the light launching setups are typically designed for achieving such a behavior. Alternatively, one can identify the use of large-core multimode hollow-core fibers as a promising avenue for the development of new photonic devices. Thus, in this manuscript, we propose and demonstrate the utilization of a large-core tubular-lattice fiber for accomplishing a speckle-based displacement sensor, which has been built up by inserting and suitably dislocating a single-mode fiber inside the void core of the hollow fiber. The work reported herein encompasses both simulation and experimental studies on the evolution of the multimode intensity distributions within the device as well as the demonstration of a displacement sensor with an estimated resolution of 0.7 µm. We understand that this investigation identifies a new opportunity for the employment of large-core hollow fibers within the sensing framework hence widening the gamut of applications of this family of fibers.**

## 1. INTRODUCTION

The most recent endeavors by the hollow-core photonic crystal fiber (HCPCF) community have entailed noteworthy advancement and led to several breakthroughs in the performance of this family of optical fibers. Indeed, the current state-of-the-art HCPCFs stand out as the lowest loss fiber optics in the visible and ultraviolet ranges [1], with attenuation figures remarkably lower than silica-core fibers, and their loss in the infrared range approaches the values corresponding to their solid-core counterparts [2]. Such a scenario grants a bright framework for HCPCF applications, which encompasses the development of optical sources [3, 4], sensors [5-7], and, most recently, power-over-fiber (PoF) platforms [8].

In the most typical HCPCF applications, the fibers and the corresponding experimental setups are designed so that the transmission and delivery of light occur in an effective single-mode fashion. Strictly speaking, the ubiquitous presence of higher-order modes (HOM) in HCPCFs' modal content forbids these waveguides from displaying a truly single-mode behavior. However, HCPCFs can be designed so that their microstructure can filter out selected families of HOMs that might be present in the fibers' modal content, hence allowing the development of fibers with very high HOM extinction ratios and capable of exhibiting effective single-mode operation from a practical outlook [9, 10]. Alternatively, HCPCFs can be designed to provide increased leakage of the core fundamental mode while keeping low loss for selected HOMs, which allows, for example, to attain fibers that favor the propagation of HOMs to the detriment of the core fundamental mode [11].

In turn, large-core HCPCFs (herein referring to fibers with core diameters larger than 100 µm) emerge as great platforms for high-power beam handling [12] and plasma photonics applications [13, 14], for example. Habitually, in beam delivery applications, once again single-mode operation is typically preferred, and the multimode characteristics of the fiber are circumvented by designing a suitable light launching setup able to entail a fundamental mode-dominated modal content. In plasma photonics applications, large-core HCPCFs are used for facilitating the ignition of microwave plasmas, which enables the emission of visible and ultraviolet light and the development of HCPCF-based fluorescent light sources [13, 14].

In the above-mentioned large-core HCPCF applications, therefore, the multimode characteristics of the fibers are either avoided or not fully explored. Delving into the multimode characteristics of large-core HCPCFs, on the other hand, has the potential of unfolding a set of exciting application possibilities from nonlinear optics to sensing. Still, investigations on the use of the multimode behavior of large-core HCPCFs are somehow scarce in the literature, mostly represented by a small collection of theoretical and fiber fabrication reports [15-18].

In this context, in this paper, we propose and demonstrate a new application of large-core HCPCFs within the sensing framework. Here, by exploring the speckled output intensity profile of a 140 μm-core diameter single-ring tubular-lattice (SR-TL) HCPCF, we were able to demonstrate a displacement sensor displaying a 0.7 μm resolution. The sensing platform is built by inserting a solid-core single-mode fiber (SMF) inside the void center of the HCPCF and assessing the speckled output profile of the latter while dislocating the SMF inside of it. In addition to the sensing results, we provide simulations and an experimental characterization of the evolution of the speckle profile along the HCPCF, as well as identify the salient features of the resulting intensity profiles. We believe that our work, by proposing and demonstrating a new application avenue for large-core HCPCFs, expands the range of HCPCF application possibilities and, especially, identifies a promising route for the exploration of the multimode behavior of HCPCFs within the sensing area.

## 2. EXPLORING THE SPECKLED OUTPUT PROFILE OF A LARGE-CORE SR-TL HCPCF

This paper investigates a displacement sensor based on the analysis of the speckled output of a large-core SR-TL HCPCF. Speckle patterns (specklegrams) appear from the interference of several modes in a multimode fiber (MMF) when excited by a coherent source. The spatial distribution of light speckles depends on parameters such as the numerical aperture of the fiber and the laser wavelength. In turn, disturbances imposed by physical and chemical stimuli can modulate the output speckle profile through mode coupling and phase deviation effects [19]. Although such intensity fluctuations are undesirable in typical optical systems, the specklegrams retain detailed information on the optical fields traveling through the fiber [20]. Therefore, one may analyze the subtle changes in the speckle output images to retrieve the characteristics of a measured variable [21].

Thus, fiber specklegram sensors (FSS) feature as feasible alternatives to the widespread intensity and FBG-based fiber sensors. Besides their high sensitivity and distributed measurement capability, FSS require a straightforward setup typically comprising a visible laser and camera instead of expensive spectrometers and optical interrogators [21, 22]. Typical FSS exploit the speckle patterns generated by conventional silica and polymer MMFs, which enables the assessment of strain [23], pressure [24], temperature [25], and refractive index [26]. Additionally, specialty fibers with holey structures provide further perspectives for modulating the output specklegram. Examples comprise exposed-core multimode fibers for temperature and chemical measurements [27, 28], and a biodegradable structured fiber made of agar suitable for biochemical and environmental assessments [29]. To our knowledge, however, the speckled output profile of large-core SR-TL HCPCFs has not been explored for sensing purposes previously.

We start our analyses by presenting simulation results on the evolution of the speckle intensity profile in a large-core HCPCF. In the experiments to be described in the following, light is launched in the HCPCF by using a SMF placed inside the HCPCF. Thus, here we approach the realization of numerical simulations (based on the beam propagation method) using the BeamPROP software (RSoft Design) to study the evolution of the optical fields emanating from the SMF until a speckled intensity distribution is established in the HCPCF. Fig. 1a shows a diagram of the 12-tube SR-TL HCPCF we study herein. Its structure comprises a set of untouching 12 tubes that defines the fiber hollow core. In the simulations, we assume that the diameters of the fiber core and cladding tubes are $D_{core}$ = 140 μm and $D_{tubes}$ = 35 μm, respectively. The thickness of the cladding tubes has been set as 680 nm.

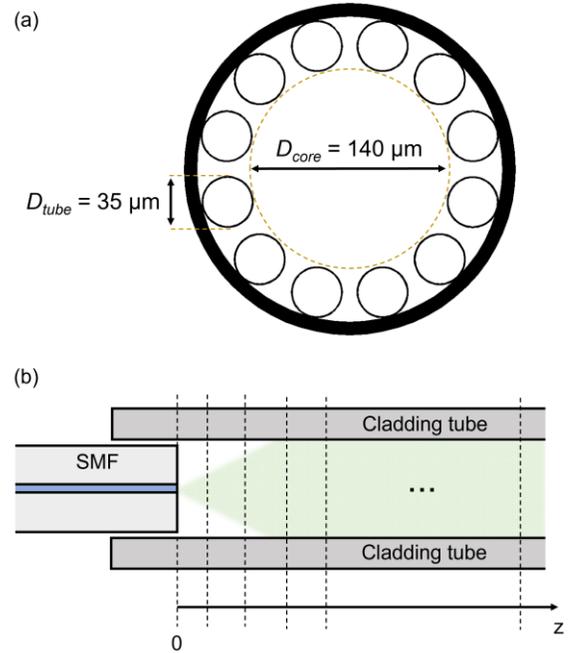

**Fig. 1.** (a) Diagram of the 12-tube SR-TL HCPCF; $D_{core}$: diameter of the core; $D_{tube}$: diameter of the cladding tubes. (b) Lateral view of the numerical setup. A SMF launches a gaussian beam in the core of the HCPCF (green shadowed region) and the intensity distributions on the HCPCF cross-section are assessed at selected positions along the fiber axis (vertical dashed lines along the z-axis).

Fig. 1b presents a lateral view of the numerical setup we employ in our analysis. A SMF with a 4 μm core diameter launches a gaussian beam (at the wavelength of 543 nm) in the HCPCF core, as represented by the green shadowed region in Fig. 1b. Such a numerical configuration allows assessing the intensity profiles on the HCPCF cross-section at selected positions along the fiber axis (vertical dashed lines along the z-axis) and, hence, study the dynamics of the formation of the speckled intensity distributions in the core of the HCPCF.

Fig. 2 displays the simulated normalized intensity distributions (normalization taken at every image, assuming as unit the maximum intensity value) in the HCPCF core at selected positions along the z-axis. Here, we define z = 0 as the position of the SMF end-face, as represented in Fig. 1b. By observing the images in Fig. 2, one sees that the mode-field-diameter of the gaussian beam emanating from the SMF enlarges as it propagates inside the HCPCF (see the images corresponding to z = 100 µm, z = 200 µm, and z = 300 µm). Additionally, one observes that, as the beam propagates, coupling to several HOMs in the hollow core occurs and, as a result, a spatial distribution stemming from the interference of the latter is developed. After hundreds of micrometers-propagation length, a speckle intensity distribution is established in the HCPCF cross-section (see the image corresponding to z = 700 µm and the consecutive ones). As one will discuss in the following, evaluation of the correlation between the speckle profiles for different propagation distances allowed us to demonstrate the system depicted in Fig. 1b as a novel platform for displacement sensing. Before presenting such results, however, we provide an experimental characterization of the formation of the speckle profile in the HCPCF.

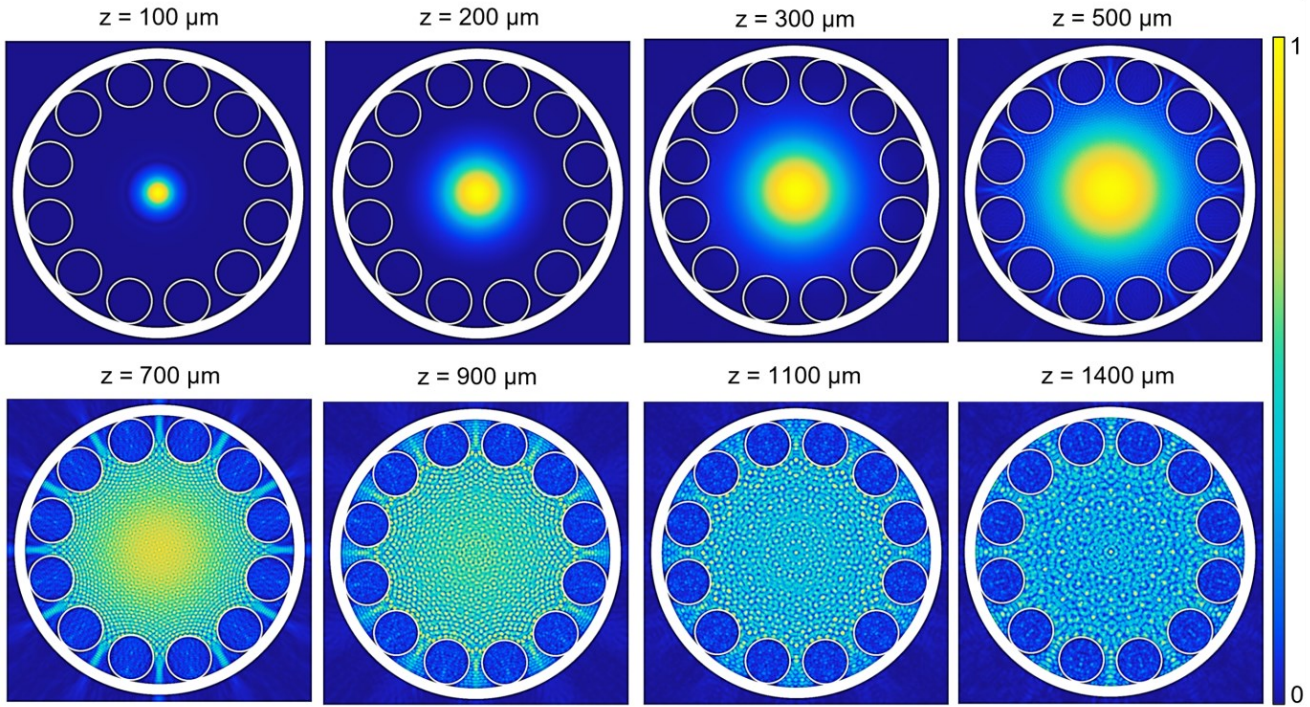

**Fig. 2.** Simulated evolution of the intensity distribution profile on the HCPCF cross-section along the z-axis at 543 nm. The color bar stands for the normalized intensity calculated for every image. Here, we define z = 0 as the position of the SMF extremity, as depicted in Fig. 1b.

Fig. 3a shows the cross-section of the SR-TL HCPCF we used in our experiments, which has been fabricated by following typical stack-and-draw procedure. The fiber displays a 140 µm-diameter core and a cladding formed by 12 tubes with a diameter of 35 µm and a thickness of 680 nm. Fig. 3b presents the transmission spectrum of such a fiber, taken for a 4 m-long fiber piece. To account for the spectrum shown in Fig. 3b, light from a supercontinuum source has been coupled to the HCPCF, and the transmitted power has been measured by an optical spectrum analyzer. Fig. 3b readily shows that the wavelength used in our characterization and sensing experiments, 543 nm, lies within the fiber second-order transmission band. The wavelength of 543 nm has been identified as a vertical dashed line in Fig. 3b.

To experimentally characterize the formation of the speckle intensity profile at the HCPCF output, we designed an experiment as represented in Fig. 4. By suitably using the alignment mirrors M1 and M2, and the coupling objective lens L1, light from a laser source emitting at 543 nm has been coupled to a SMF (Newport F-SA, 125 µm outer diameter), which, in turn, had been previously and carefully inserted in the core of the HCPCF with a length of 5 cm. The SMF has been, in sequence, attached to a motorized stage (step resolution of 0.1 µm) so it could be displaced inside the HCPCF.

At the output of the HCPCF, one placed an objective lens (L2) for collimating the fiber output beam and projecting it onto a DSLR camera (Sony Alpha 100 DSLR). We remark that we removed the DSLR camera lens so that the output beam could directly impinge on the camera sensor. In the experiments, differently from the simulations, we redefined z = 0 at the output of the HCPCF and changed the orientation of the z-axis, as illustrated in Fig. 4. Such an alteration of the z = 0 definition and axis orientation with respect to the simulations allows us to directly correlate the simulated and measured output intensity distributions, as the comparisons rely on the propagation lengths of the optical fields inside the HCPCF.

Thus, with the aid of the motorized stage, the SMF could be controllably drawn from the HCPCF starting from the newly defined z = 0 position (HCPCF output end face, see Fig. 4). The initial positioning of the SMF has been assured by tuning the focal plane of the objective lens L2 to be at the HCPCF end face as it allowed, by suitably moving the SMF inside the HCPCF, to adjust its position so that the SMF and HCPCF extremities were at the same plane. While drawing the SMF from the HCPCF, images of the beam emerging from the HCPCF could be captured by using the DSLR camera at each translation step. It allowed us to qualitatively reproduce the simulated images shown in Fig. 2, as one will see in the following.

Fig. 5 presents the images captured by the DSLR camera at representative positions of the SMF output along the z-axis (a schematic of the cladding tubes has been superposed to the images for eye guidance). As in Fig. 2, we observe the enlargement of the mode-field-diameter of the gaussian beam emanating from the SMF as it propagates inside the HCPCF (as one can observe in the images corresponding to z = 100 µm, z = 200 µm, and z = 300 µm). Additionally, and consistently with the simulations shown in Fig. 2, one sees that coupling to HOMs in the HCPCF takes place after hundreds of micrometers of propagation length and that a multimode intensity distribution is attained in the hollow core. One notice that the intensity distributions in the images shown in Fig. 5, especially for the images corresponding to z ≤ 500 µm, are slightly shifted from the HCPCF geometrical center. While it is likely to be due to misalignment of the SMF inside the HCPCF, as the inner diameter of the HCPCF core (140 µm) is slightly greater than the outer diameter of the SMF (125 µm), it does not prevent us to conclude that our simulations and experiments are qualitatively consistent and, hence, that we could characterize the evolution of the gaussian beam emanating from the SMF up to the establishment of a multimode intensity profile in the HCPCF.

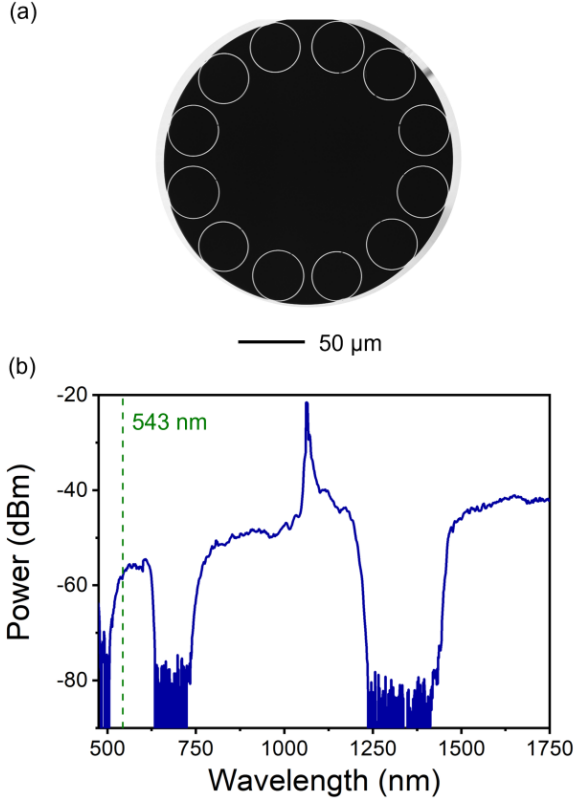

**Fig. 3.** (a) Cross-section of the 12-tube SR-TL HCPCF. (b) Transmission spectrum for a 4 m-long fiber. The vertical dashed line identifies the wavelength at which the characterization and sensing experiments have been performed, 543 nm.

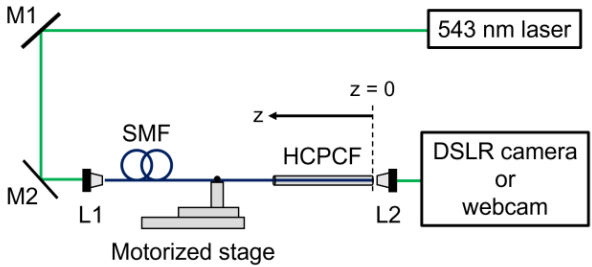

**Fig. 4.** Diagram of the experimental setup for characterization and sensing measurements. The SMF (in blue) is carefully inserted in the HCPCF void core and displaced towards +z direction during the measurements. M1 and M2: alignment mirrors. L1 and L2: objective lenses.

## 3. FIBER SPECKLEGRAM ANALYSIS AND DISPLACEMENT SENSING

The intensity distribution of an output speckle pattern $I(x, y)$ projected on the xy-plane can be described by Eq. (1):

$$I(x,y) = \sum_{m=1}^{M} \sum_{n=1}^{M} a_m a_n \exp[i(\phi_m - \phi_n)], \quad (1)$$

where $M$ is the number of modes, $a(x, y)$ and $\phi(x, y)$ are the amplitude and phase distributions of the optical modes, and the subscripts denote the $m$-th and $n$-th fiber modes [21]. Let $I_0(x, y)$ be the specklegram for the initial fiber state. Disturbances introduced by fiber bending, temperature changes, laser power fluctuations, and deviations in the launching conditions modulate the speckle pattern and result in another state $I$ contrasting to $I_0$. The zero-mean normalized cross-correlation (ZNCC) metric quantifies the differences between $I$ and $I_0$ according to Eq. (2) [30]:

$$Z = \frac{\iint (I - \bar{I})(I_0 - \bar{I}_0) dx dy}{[\iint (I - \bar{I})^2 dx dy \iint (I_0 - \bar{I}_0)^2 dx dy]^{1/2}}, \quad (2)$$

where $0 \leq Z \leq 1$ is the ZNCC coefficient and the overlines indicate the average values of $I$ and $I_0$. Therefore, one sees that $Z = 1$ identifies the situation in which the fiber state matches the reference condition ($I = I_0$) and that the Z value decreases as $I$ departs from $I_0$.

Despite being sensitive and intrinsically robust to illumination changes, the ZNCC exhibits an impaired measurement range due to its saturation. Thus, one may compute the extended ZNCC (EZNCC) to overcome such limitation according to Eq. (3) [31]:

$$EZ = EZ_0 - [Z(I, I_0, \tau) - 1], \quad (3)$$

where $EZ$ is the EZNCC coefficient, $EZ_0$ is a reference value, $Z$ is the ZNCC between $I$ and $I_0$, and $\tau = 0.7$ is an empiric threshold parameter. The algorithm starts with $EZ_0 = 1$, i.e., $EZ = 1$ for the undisturbed condition ($I = I_0, Z = 1$). The evaluation hence proceeds by decreasing $EZ$ until $Z$ reaches $\tau$. This condition resets the reference $I_0$ to the current specklegram $I$ and makes $EZ_0 = EZ$, preventing its saturation. Consequently, the EZNCC exhibits linear response for a virtually unlimited dynamic range [31]. Thus, here we have chosen the EZNCC as the metric for analyzing our sensing data.

By following the methods described above, and by using the setup depicted in Fig. 4, we could attain displacement sensing measurements via the assessment of the correlation values as a function of the position of the SMF inside the HCPCF. We remark that, during the displacement sensing measurements, the DSLR camera has been replaced by a low-cost commercial lensless webcam (800 × 600 pixels). Using a webcam instead of a DSLR camera allows us to both attain a cost-effective sensing platform as well as to readily recover and analyze the recorded images with a computer using MATLAB (Mathworks) routines.

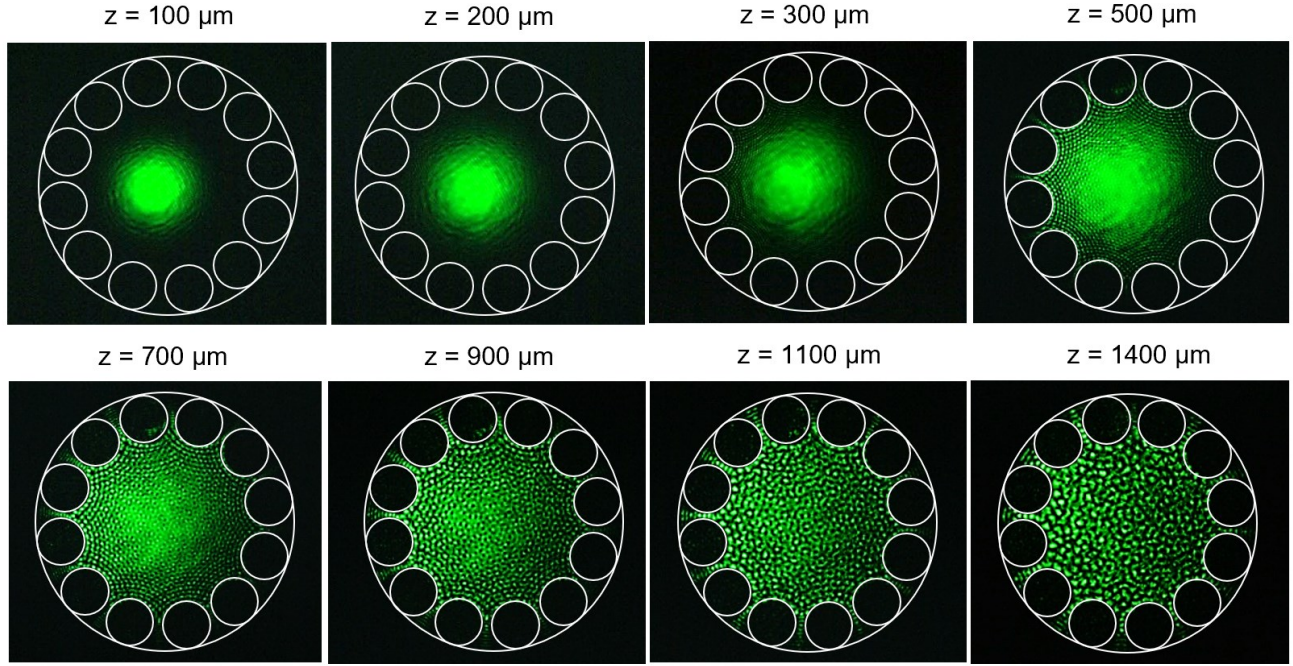

**Fig. 5.** Experimental observation of the evolution of the intensity distribution profile at 543 nm on the HCPCF cross-section as a function of the position of the SMF inside of it. The images have been taken by using a DSLR camera, as depicted in Fig. 4. The tubes' location on the fiber cross-section have been superposed to the captured images for eye guidance.

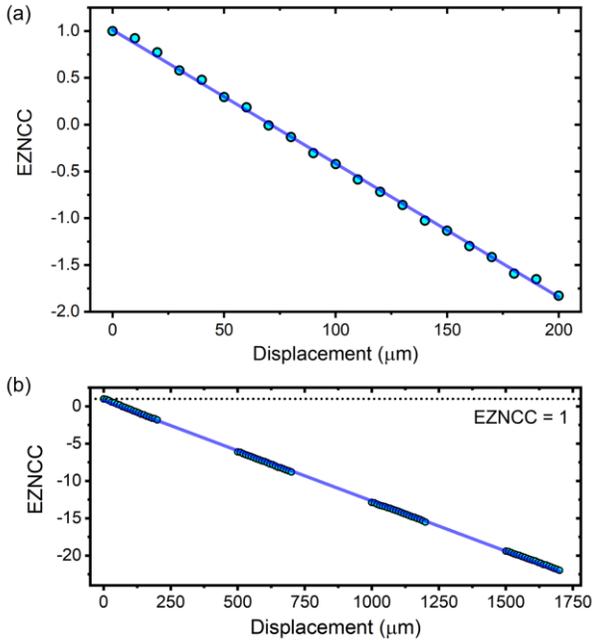

**Fig. 6.** Displacement sensing results. Plots of the EZNCC values as a function of the SMF displacement inside the HCPCF for a measurement range of (a) 200 μm and for an extended range of (b) 1.7 mm.

The sensing results are shown in Fig. 6, where the EZNCC values have been plotted as a function of the displacement of the SMF inside the HCPCF. We mention that, in the sensing measurements, the initial position of the SMF has been set to be 3 mm apart from the HCPCF extremity and that the SMF movement has been directed towards the positive direction of the z-axis (see Fig. 4). Such a choice of the departure point of the SMF in the sensing measurements considered that, according to our observations, it was the position at which the speckled profile at the HCPCF output was reasonably developed and adequate for sensing purposes.

Fig. 6a shows the results of the displacement sensing test taken for a measurement range of 200 μm. Each data point summarizes the average of 70 specklegram images and considers twice the standard deviation of the mean ($2\sigma$) as their error bar – as the $2\sigma$ value amounts to 0.03 at maximum, the error bars are not visible in the plots in Fig. 6. Observation of the data allows identifying a linear trend between the EZNCC and displacement values. Thus, by linear-fitting the data, we could estimate the device sensitivity as $14.27 \pm 0.07$ mm$^{-1}$ ($R^2 = 0.9996$). To demonstrate that our sensing setup can have an extended measurement range, we performed additional measurements for displacements between 500 μm and 700 μm, 1000 μm and 1200 μm, and 1500 μm and 1700 μm. The results are shown in Fig. 6b, where one can observe that the linear behavior of the EZNCC as a function of the displacement is maintained. For the extended-range measurement shown in Fig. 6b, a sensitivity of $13.47 \pm 0.02$ mm$^{-1}$ ($R^2 = 0.9997$) has been calculated.

Finally, given the maximum random uncertainty of 0.01 (in EZNCC units), one can estimate the system resolution as $\Delta EZ/\Delta z = 0.01/13.47 \approx 0.7$ μm. Such a resolution figure compares well with the ones estimated in other fiber optics platforms as fiber Bragg gratings and multicore fiber-based displacement sensors [32, 33]. Furthermore, the EZNCC algorithm enlarges the measurement range beyond typical fiber speckle-based strain and displacement meters [23, 34] by preserving its intrinsic high-sensitivity feature.

Moreover, we consider that the platform we reported in this paper can disclose a set of other speckle-based sensing applications using large-core HCPCFs. One can envisage, for example, that our approach could also be applied to the monitoring of other parameters such as gas pressure levels, as one could expect that the alteration of the gas pressure inside the HCPCF would impact its output speckle pattern. Additionally, further studies can encompass the investigation of the temperature sensitivity of the devices since the lower thermo-optic coefficient of air compared to silica can entail lessened temperature cross-sensitivity and, thus, be beneficial to practical devices. Furthermore, we consider that our approach could be potentially useful in applications using liquid-filled HCPCFs, hence providing interesting avenues for chemical and biological sensing.

## 4. CONCLUSIONS

In this manuscript, we proposed and demonstrated a novel sensing application using a large-core HCPCF. Thus, by assessing the speckled output intensity distribution of a 140 μm-core diameter SR-TL HCPCF, we could demonstrate a displacement sensor displaying a 0.7 μm resolution. The sensing platform reported herein has been built by introducing a solid-core SMF in the void center of the HCPCF and evaluating the speckled output profile of the HCPCF while moving the SMF inside of it. Additionally, we provided simulations and an experimental characterization of the evolution of the intensity distributions inside the HCPCF starting from the gaussian beam emanating from the SMF up to the establishment of the multimode intensity distribution in the HCPCF. We consider that our work, by proposing and demonstrating a new application of large-core HCPCFs, enlarges the range of HCPCF application opportunities and, particularly, discloses a promising path for the study of the multimode behavior of HCPCFs in sensing.

**Funding.** Fundação de Amparo à Pesquisa do Estado de São Paulo (FAPESP, grants 2021/13097-9 and 2017/25666-2); Conselho Nacional de Desenvolvimento Científico e Tecnológico (grants 403418/2021-6, 310650/2020-8 and 309989/2021-3).

**Disclosures.** The authors declare that there are no conflicts of interests related to this article.

**Data availability.** Data underlying the results presented in this paper are available from the corresponding author upon reasonable request.